\documentclass[aps, prl, letterpaper, 10pt, twocolumn]{revtex4-1}%superscriptaddress
\usepackage[colorlinks=true,allcolors=green!55!blue]{hyperref}
\usepackage{amssymb}
\usepackage{mathptmx}
\usepackage{amsmath}
\usepackage{graphicx}
\usepackage{subfigure}
\usepackage{amsfonts}
\usepackage{xcolor}
\usepackage{epsfig}
\usepackage{color}
\usepackage{bbm}
\usepackage{bm}
\usepackage{tabularx}
\usepackage{multirow}
\usepackage{mathtools}
\usepackage{xfrac}
\DeclareMathAlphabet{\mathcal}{OMS}{cmsy}{m}{n}

\newcommand{\ket}[1]{\vert{#1}\rangle} 
\newcommand{\bra}[1]{\langle{#1}\vert}
\newcommand{\proj}[1]{\ket{#1}\!\bra{#1}}

\newcommand{\hatd}[1]{\hat{#1}^{\dagger}}

\newcommand{\gele}{\textsl{g}_{\rm e}}

\begin{document}
\title{Dynamical Topological Phase Transition in a Bose-Einstein Condensate}

\author{Mehdi Abdi}
\email{mehabdi@gmail.com}
\affiliation{Department of Physics, Isfahan University of Technology, Isfahan 84156-83111, Iran}

\begin{abstract}
We study a Bose-Einstein condensate (BEC) at low energy limit and show that their collective dynamics exhibit interesting topological behavior. The system undergoes dynamical topological phase transition at its global periods if its dispersion relation is strictly linear and when all of the bosonic mode are properly displaced from their equilibrium position.
We corroborate the occurrence of dynamical phase transition by calculating Fisher zeros of the Loschmidt amplitude for sudden quench dynamics.
A connection is established between the order of nonanalycity in the accumulated geometric phase and the spectral density of the system.
Furthermore, it is shown that a power law scaling holds at all critical times for various displacement spectra, whose dynamical exponent equals unity.
Eventually, a scheme for the quantum simulation of such dynamical phase transition is proposed.
The scheme is based on the vibrational spectrum of a free-standing membrane of a two-dimensional material.
To induce a displacement in the system and for tracking its collective geometrical dynamics, we propose to employ a spin that properly couples to the modes.
When appropriate geometrical and boundary conditions are applied to the membrane, a spectrum with linear dispersion is attainable, then decoherence dynamics of the spin unveils occurrence of a dynamical topological phase transition.
\end{abstract}

\maketitle

%
%
%========================================%
\textit{Introduction.---}%
Quantum phase transitions (QPTs) have always been offering attracting physics~\cite{Vojta2003, Sachdev2011}. Dynamical behavior of quantum systems far from equilibrium can also exhibit abnormalities in the physical observables; signaling a dynamical QPT~\cite{Zvyagin2016, Heyl2018}.
The dynamical QPT is an emerging topic expanding the concept of phase transition to the seemingly uncontrollable parameter, the time, and lies at the heart of sudden quenchs~\cite{Sengupta2004, Zurek2005, Kollath2007, Braun2015}.
Alongside, dynamical topological phase transitions (DTPT) in many body systems have become growingly studied because of their importance in addressing fundamental questions~\cite{Vajna2015, Dong2015, Wang2015, Flaeschner2017, Zhou2018, Wang2019}.
The occurrence of DTPT in a quantum system is not necessarily associated with its equilibrium topological properties that are conventionally addressed by the topological invariants~\cite{Schnyder2008}. In fact, the DTPTs can even occur `accidentally' when the initial and final quench Hamiltonians are topologically equivalent~\cite{Vajna2015}.
On the other hand, a correspondence between the ground state geometric phase, as a topological feature, and criticality of the Ising model and a central spin coupled to such systems is established~\cite{Zhu2006, Yuan2007}, revealing the connection of these two supposedly irrelevant aspects.
The quantum phase transitions has always been associated with nonlinearity of the system. For this, the spin and fermionic systems such as Ising, Heisenberg, and two-band models have been theoretically predicted to exhibit the effect~\cite{Diehl2010, Heyl2013, Huang2016, Halimeh2017, Zunkovic2018, Jafari2019} and are even experimentally confirmed~\cite{Jurcevic2017}.
Instead, in the case of bosons a highly nonlinear interacting systems, namely the Bose-Hubbard model, have been shown to have the grounds for undergoing a QPT~\cite{Cazalilla2011, Lacki2019}.
Nonetheless, the occurrence of these dynamical QPTs is not always linked to the equilibrium state of the system~\cite{Hickey2014, Zauner2017, Jafari2019}.

Bose-Einstein condensates at weakly interacting regime are described by a set of non-interacting bosons forming a linear system.
In a very recent experiment the geometric phase of a BEC is measured via a dynamical quantum Zeno effect~\cite{Do2019}.
Such a system can be brought into interaction with an atomic degree of freedom, e.g. via laser transitions. The system has shown to provide interesting physics, especially, it has a close correspondence with the spin-boson model at sufficiently low energy and temperatures~\cite{Recati2005, West2013}. When the above mentioned interaction is tuned properly, pure dephasing of the atomic state can be used as a non-destructive probe to the temporal phase fluctuations of the BEC~\cite{Bruderer2006}.

%our focus is on a linear bosonic system whose geometric phase behaves non-analytically.
In this paper, we show that despite its seemingly trivial dynamics, a linear bosonic system, experiences phase transitions owing to its topological properties:
A 1D Bose liquid that forms a system of non-interacting bosons with a linear dispersion relation undergoes a DTPT after a sudden quench to a displaced  phase. That is, quenching from the state where the global wavefunction accumulates no geometric phase to the topologically nontrivial state.
%We study the parameter regions that the effect is available and propose a setup to experimentally observe and examine it.
We show that the return amplitude behaves non-analytically stemming from the kinks in the collective geometric phase.
%It is revealed that the order of non-analicity depends on the displacement function of the modes. 
The dynamical topological order parameter (DTOP) is employed for quantifying behavior of the kinks~\cite{Budich2016}. The occurrence of a dynamical QPT is witnessed by showing that the cusps occur where the Fisher zeros of the Loschmidt amplitude cross the real time axis, the critical times $t_c$. By investigating the behavior of DTOP (and its derivatives) at the critical points we find a power law scaling regardless of the spectral density and the critical time.
Due of the correspondence between a BEC interacting with an atom and the spin-boson model we use our proposed setup for spin--boson model based on color centers in hexagonal boron nitride (h-BN) monolayers~\cite{Abdi2018a} to show such DTPT in the set of bosons manifests itself as kinks in the qubit coherence dynamics.
The spin--boson model, in turn, is realized by coupling the electronic spin of a color center to vibrations of a free-standing h-BN membrane~\cite{Abdi2018}.
\begin{figure}[tb]
\includegraphics[width=\columnwidth]{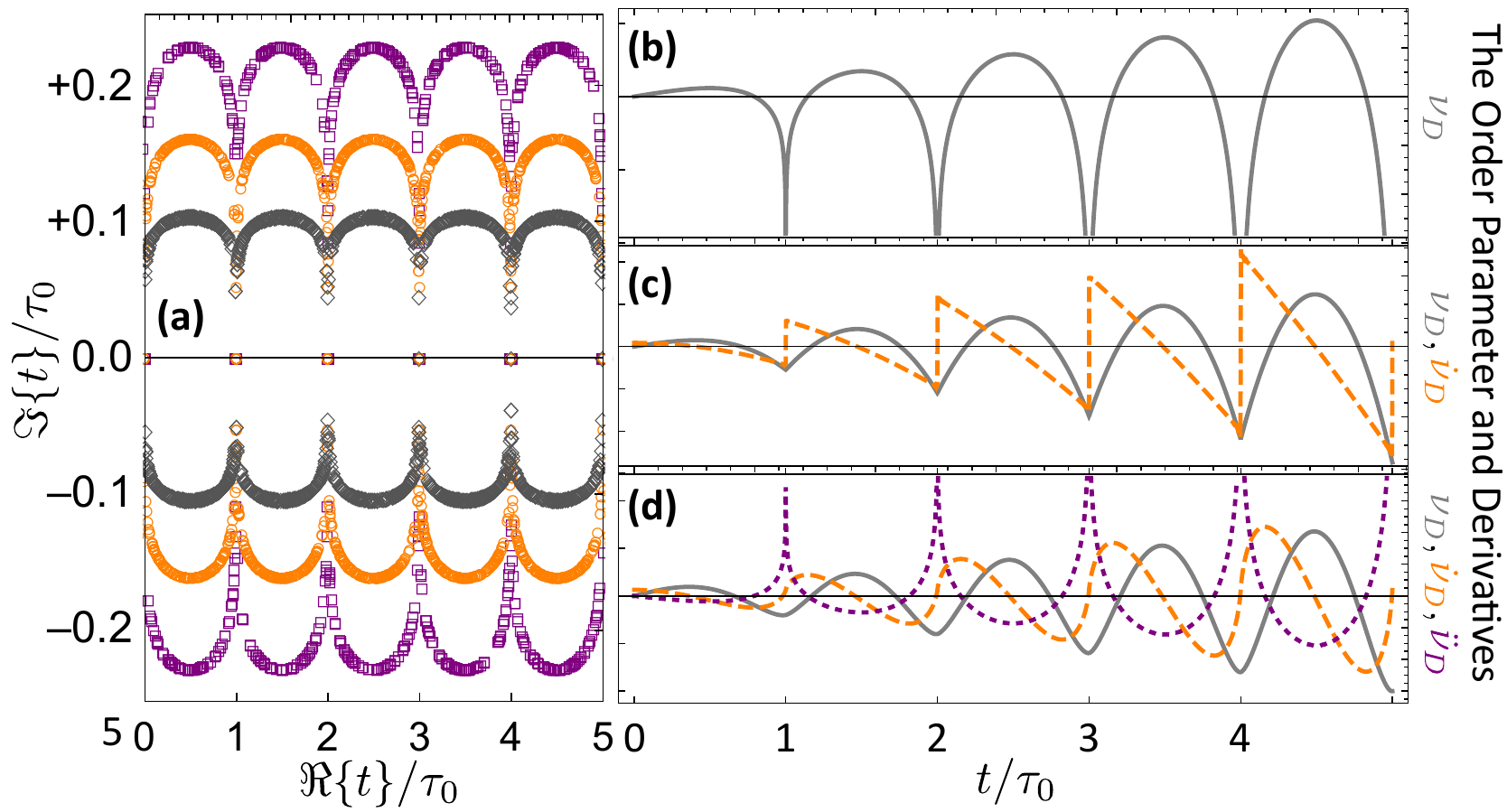}
\caption{%
(a) Fisher zeros for $\alpha=+1$ (gray), $\alpha=0$ (orange), and $\alpha=-1$ (purple).
(b)--(d) Dynamical topological order parameter $\nu_D$ (gray solid) and its first (dashed orange) and second (dotted purple) derivatives for $\alpha=+1,0,-1$, respectively.
The system size is $N=10^3$.
}
\label{fig:DPT}%
\end{figure}
%

%
%
%========================================%
\textit{Model.---}%
The Hamiltonian of an ideal boson gas interacting in pairs is given by
%\begin{equation}
$\hat H= \sum\epsilon(\mathbf{p})\hatd{a}_{\bf p} \hat{a}_{\bf p}
+U\sum\hatd{a}_{\mathbf{p}_1}\hatd{a}_{\mathbf{p}_2} \hat{a}_{\mathbf{p}_2'}\hat{a}_{\mathbf{p}_1'}$
%\end{equation}
where $\epsilon$ is the energy and $\mathbf{p}$ are momenta of the bosons, respectively~\cite{Lifshitz1980}.
The sums are over the momenta respecting the conservation of momentum, while $U$ is the scattering amplitude.
For weak scattering rates the nonlinear term is treated in mean-field approximation and a Bogoliubov transformation applied to arrive at an effective harmonic-fluid description that describes a 1D quantum fluids at low energy limits~\cite{Haldane1981, Chiao2000}.
The only available excitations at the low energies are a set of phonons with a \textit{linear dispersion relation} $\omega_{\bf p}\propto|\mathbf{p}|$.
Therefore, the dynamics of such a confined BEC reduces to a bath of non-interacting bosons with equally spaced frequencies~\cite{Pethick2002}. That is, the frequency of the $k$th mode is given by $\omega_k = (k+1)\omega_0$ for $(k=0,1,2,\cdots)$ with $\omega_0$ the lowest frequency of the system. Therefore, the fundamental mode frequency literally determines the period of the whole system $\tau_0 = 2\pi/\omega_0$.
%Note that we have substituted a continuous with a discretized momentum space. Furthermore, in the momentum space the system is one-dimensional and possesses a linear dispersion relation. Therefore, momentum value of the modes $k$ are indeed quantized suggesting that the system is bounded.
Here, we emphasize that the linear dispersion relation is at the heart of the observations that are discussed in this paper. In fact, the DTPT could not occur for nonlinear or even partly linear systems~\cite{suppinfo}.

The Hamiltonian of such system, apart from an irrelevant constant, reads $\hat H_0 = \sum_{k=0}^N\omega_k\hatd b_k\hat b_k$ where $N$ is size of the system (with the assumption of a bounded system). The ground state is $\ket{\emptyset}=\bigotimes_k\ket{0}_k$ whose time evolution is trivial and only collects a global dynamical phase corresponding to the vacuum energies.
Nevertheless, when the modes are locally displaced by $\hat V_{k}=g_k(\hat b_k +\hatd b_k)$ the ground state turns into a multimode coherent state $\bigotimes_k\ket{\alpha_k}$ with amplitudes $\alpha_k = 2g_k/\omega_k$. In the phase space representation each mode is associated to a circular bulb satisfying the minimum Heisenberg uncertainty condition in the respective phase $x_kp_k$ plane. Time evolution of the system now becomes nontrivial as each bulb goes round a circle in their phase plane passing through the origin and centered at $(x_k,p_k)=(g_k/\omega_k,0)$.
The modular variables of each mode are defined by $\hat{b}_k=(\hat{x}_k+i\hat{p}_k)/\sqrt{2}$ and hence follow the algebra with all of their commutators equal to zero but $[\hat{x}_j,\hat{p}_k]=i\delta_{jk}$.
Due to the non-commuting nature of the modular variables each mode accumulates a geometric phase as it evolves by time, which is proportional to the area enclosed by its circular path: $\phi_k(t)=(g_k/\omega_k)^2[\omega_k t -\sin\omega_k t]$~\cite{Asadian2014}.
The total geometric phase is then
\begin{equation}
\Phi_{\rm G}(t)=\sum_{k=0}^N(\frac{g_k}{\omega_k})^2\big[\omega_k t -\sin\omega_k t\big],
\label{geo}
\end{equation}
which is composed of a linear term in time and an oscillating part. For the sake of clarity, we discard the effect of linear term in the following as it does not contribute in the phase transitions.
In the cases where the $g_k$ values are chosen properly the total geometric phase or its derivatives exhibit non-analytic behaviors~\cite{suppinfo}.
The kinks in the geometric phase, in turn, signal criticality of the system~\cite{Carollo2005, Zhu2006, Hamma2006} and---as it will become clear shortly---they indeed refer to a \textit{dynamical topological phase transition}~\cite{Heyl2013, Zunkovic2018, Flaeschner2017}. The concept which is studied in the rest of paper.
The order of criticality of $\Phi_{\rm G}$ is determined by the displacement parameters. Assuming $g_k \propto \omega_k^{\alpha/2}$, it is useful to instead work with the spectral density function introduced as $\mathcal{J}(\nu) = \sum_k (g_k/\omega_k)^2 \delta(\nu -\omega_k)$. The linear dispersion relation of our system simplifies the function to $\mathcal{J}(\nu) \propto \nu^{\alpha}$ in the continuum limit. 
For values about $\alpha = 1$ the geometric phase of the BEC system suffers zeroth order criticality at the global periods $\tau_0$. Instead, it experiences a first order criticality (kinks in the first derivative of total geometric phase $\dot\Phi_{\rm G}$) around $\alpha = 0$, and so on such that for a given value of $\alpha$ a $(1-\alpha)$th criticality is expected with respect to the total geometric phase.
Interestingly, the effect disappears as soon as the dispersion relation deviates from linear a function~\cite{suppinfo}.

%
%
%========================================%
\textit{Dynamical topological phase transition.---}%
The critical behavior of the total geometric phase basically manifests itself as kinks in the Loschmidt amplitude of the quench dynamics in the system. That is, when the system is prepared in the ground state of $\hat H_0$ and then rapidly enforced to follow the dynamics dictated by $\hat H_1=\hat H_0 +\sum_k \hat V_k$ there is a finite probability of return to the ground state. The amplitude of such return probabilities, the so-called Loschmidt amplitude, is given by 
\begin{equation}
\mathcal{G}(t)\equiv \bra{\emptyset}\exp\!\{-it\hat{H}_1 \big\}\ket{\emptyset}.
\label{loschmidt}
\end{equation}
This quantity is the dynamical counterpart of the statistical boundary partition function and its roots in the complex time plane are called the Fisher zeros.
A dynamical phase transition occurs where locus of the zeroes of $\mathcal{G}(z)$ in the complex time plane cross the real axis~\cite{Heyl2013}.
An analytical expression is found for the Loschmidt amplitude in our system $\mathcal{G}(z)=\exp\{-\sum_{k=0}^N(g_k/\omega_k)^2(1-\cos\omega_k z)\}$~\cite{suppinfo}.
We then numerically find the complex zeros for $\alpha=0, \pm 1$ and notice they crossing the real time axis at $t_n=n\tau_0$ for $n=1,2,3,\cdots$ [Fig.~\ref{fig:DPT}(a)].
\begin{figure}[tb]
\includegraphics[width=\columnwidth]{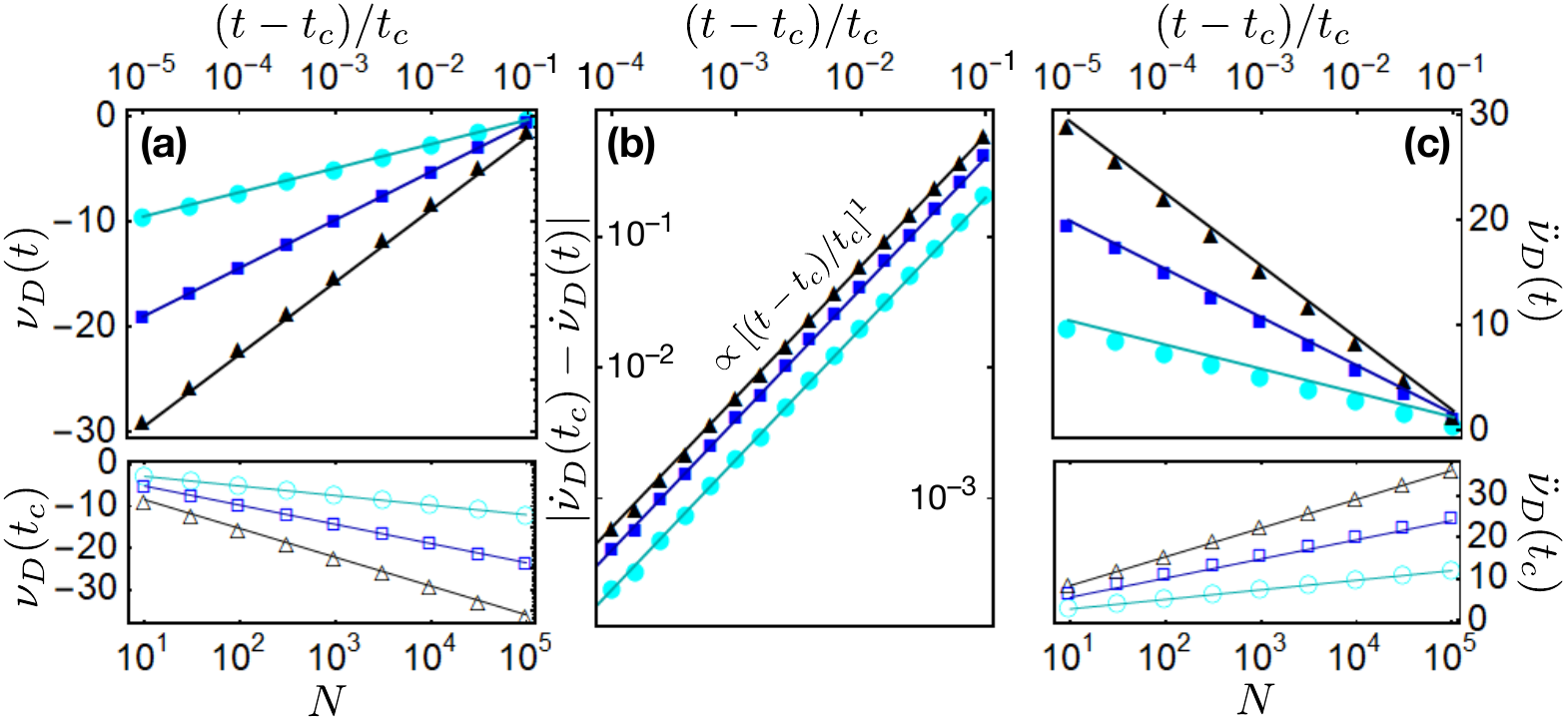}
\caption{%
Scaling of $\nu_D$ and its derivatives for $\alpha=+1,0,-1$ from (a) to (c), respectively.
In (a) and (c) logarithmic scaling of the order parameter is shown by fitting to Eqs.~(\ref{scal1}) and (\ref{scal2}), while in (b) the power law scaling for the three first critical times is presented.
In all plots $t_c=t_1$ (cyan), $t_c=t_2$ (blue), and $t_c=t_3$ (black).
The system size is $N=10^5$.
}
\label{fig:scale}%
\end{figure}
The topological nature of the transitions is further revealed by following the variations of the dynamical topological order parameter (DTOP) introduced in~\cite{Budich2016, Bhattacharya2017}:
%Here, due to the linear dispersion of the system we instead use an expanded definition of the parameter by summing over the whole momentum space:
\begin{equation}
	\nu_D(t)=\sum_{k=0}^N \frac{\partial\phi_k(t)}{\partial \omega_k}.
\end{equation}
Therefore, the DTOP is no longer limited to integer values and, in general, is a smooth function of time except at the critical points that are determined by the Fisher zero lines.
In Fig.~\ref{fig:DPT} this parameter is plotted for the three different spectral densities. The order of criticality follows the pattern described for the geometric phase. That is, for integer values of $\alpha$ the system experiences a $(1-\alpha)$th dynamical topological phase transition~\cite{Canovi2014, Vajna2015, Flaeschner2017}.
The Fisher zeros corresponding to these systems are depicted in Fig.~\ref{fig:DPT}(a). The lines that are formed by the roots cross the real time axis at integer multiples of the global system period $t_n=n\tau_0$ ($n=0,1,2,\cdots$) confirming the occurrence of the phase transition.
More interestingly, they show that as the value of $\alpha$ decreases the density of Fisher zeros is decreased and they distance from the real time axis. This explains the order of DTPT for each case.
In general, two types of critical behavior is noticed  from Fig.~\ref{fig:DPT}: (i) The logarithmic divergences at the critical times $t_c$ that arise for systems with spectral exponent $\alpha=\pm 1$. (ii) Discontinuity of $\dot\nu_D$ occurring when $\alpha=0$.
Notwithstanding, the scaling properties of both types are found to be the same (see below).
For the criticalities of type (i) we find that the values of $\nu_D$ and $\ddot\nu_D$ at the critical points $t_c$ diverge logarithmically with increasing the system size as
\begin{equation}
f(t_c) = \kappa\log N +\upsilon,
\label{scal1}
\end{equation}
where $f$, given the case, is $\nu_D$ or $\ddot\nu_D$, while $\upsilon$ is an irrelevant constant.
On the other hand, these singularities at the vicinity of the critical times behave like
\begin{equation}
f(t) =\kappa'\log|(t-t_c)/t_c| +\upsilon'.
\label{scal2}
\end{equation}
According to the scaling ansatz in the case of logarithmic divergence~\cite{Barber1983}, the ratio $|\kappa'/\kappa|$ gives the dynamical exponent of the quantum phase transition~\cite{Sachdev2011}. The coefficients for $\alpha=\pm 1$ at the critical time $t_n$ are $\pm\kappa=\mp\kappa'=n$ [see Fig.~\ref{fig:scale}(a) and (c)]. Therefore, one calculates unity for the dynamical exponent.
The criticality of type (ii) is studied by investigating variations of the rescaled value of $\dot\nu_D$ at the critical time points. A power law scaling is found whose dynamical exponent turns out to be the same [Fig.~\ref{fig:scale}(b)].

The criticality in the collective dynamics of the system should be traced back to the geometrical correlations between the modular variables that manifest themselves in the total geometric phase.
Higher order time correlations are anticipated to further unveil the nature of phase transition~\cite{Asadian2014}.

\begin{figure}[b]
\includegraphics[width=\columnwidth]{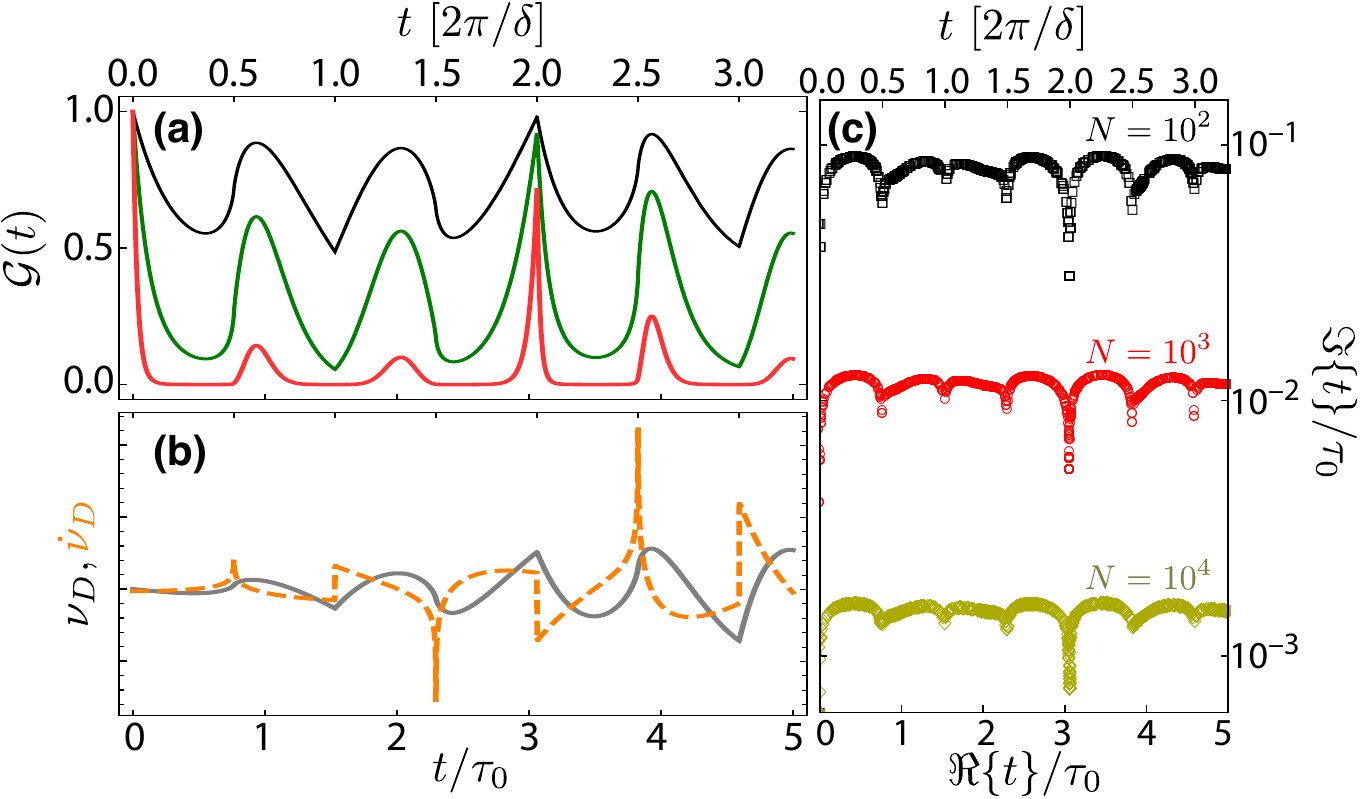}
\caption{%
(a) Collapse and revival in the spin coherence at the coupling rates: $g_0/\omega_0=0.5$ (black), $g_0/\omega_0=1.0$ (green), and $g_0/\omega_0=2.0$ (red).
(b) The dynamical topological order parameter (gray) and its derivative (orange dashed).
(c) Fisher zeros of the Loschmidt amplitude for three different system sizes. The zeros approach to the real time axis as the system size increases. The tongues suggest that the crossings occur at the integer multiples of $\pi/\delta$ in the thermodynamical limit.
}
\label{fig:fid}%
\end{figure}
%

%
%
%========================================%
\textit{Implementation.---}%
To experimentally implement a system that shows the above mentioned nontrivial dynamics, here we put forth a setup with a detection scheme.
The system is composed of a free-standing monolayer h-BN membrane with an embedded color center~\cite{Tran2015, Abdi2017, Abdi2018a}.
The latter is responsible for displacement of the modes and also is employed as a probe to the global geometric phase of the system~\cite{Yuan2007}. The color center has a spin doublet electronic ground state $\{\ket{\uparrow}, \ket{\downarrow} \}$. When immersed in a magnetic field gradient, the spin couples to the position of the membrane $\hat X$. The dynamics is described by the interaction Hamiltonian $\hat H_{\rm int} = \gele\mu_{\rm B}\eta\proj{\uparrow} \hat X$, where $\eta$ is the magnetic field gradient, while $\mu_{\rm B}$ and $\gele$ are electron Bohr magneton and g-factor, respectively~\cite{Abdi2018a}.
For the purpose of implementing the above idea we take a circular geometry: a membrane with radius $R$ and thickness $h$ with the defect at its center. Therefore, the spin only couples to the axisymmetric modes~\cite{Abdi2016}. The membrane is assumed to be subject to a dominant tensile force at the boundaries $\varepsilon \gg (h/R)^2$. The motion of the membrane is then described by its normal vibrational mode spectrum $\{\omega_i\}$ that span from the lowest frequency $\omega_0$, the fundamental mode, to the highest $\omega_N$, set by the maximum vibrational wavelength.
%, which in turn is lower-bounded by the lattice constant.
%The Hamiltonian of such a system is
%where $\hat H_{\rm m}$ is the mechanical Hamiltonian with $\hat X$ denoting position of the membrane with
%In terms of normal mechanical modes, the spin interacts with a vibrational spectrum that undergo harmonic oscillations. Hence, the system exhibits a `natural' spin--boson model~\cite{Breuer2007}.
We expand the displacement operator in terms of the normal modes $\hat X=\sum_{k=0}^N x_{k}^{\circ}(\hat b_k +\hatd b_k)$ to arrive at the total Hamiltonian
\begin{equation}
\hat H = \sum_{k=0}^{N}\omega_k\hatd b_k\hat b_k +g_k\proj{\uparrow} (b_k +\hatd b_k).
\end{equation}
Here, the bosonic annihilation (creation) operator $\hat b_k$ ($\hatd b_k$) is assigned to the $k$th mechanical normal mode. The coupling strength of each mode to the spin-qubit is given by $g_k \equiv \gele\mu_{\rm B}\eta x_{k}^{\circ}$ with the zero-point fluctuation amplitude of $k$th mode $x_{k}^{\circ}=\sqrt{\hbar/2m_k^*\omega_k}$, where $m_k^*$ is effective mass of the mode.
Such vibrational bath is found to have normal modes with equally spaced frequencies $\omega_{k+1}-\omega_k = \delta$, hence a linear dispersion relation. And all modes are equally coupled to the spin. Its spectral density is thus of white-noise with $\alpha=0$~\cite{Abdi2018a}.

\begin{figure}[tb]
\includegraphics[width=\columnwidth]{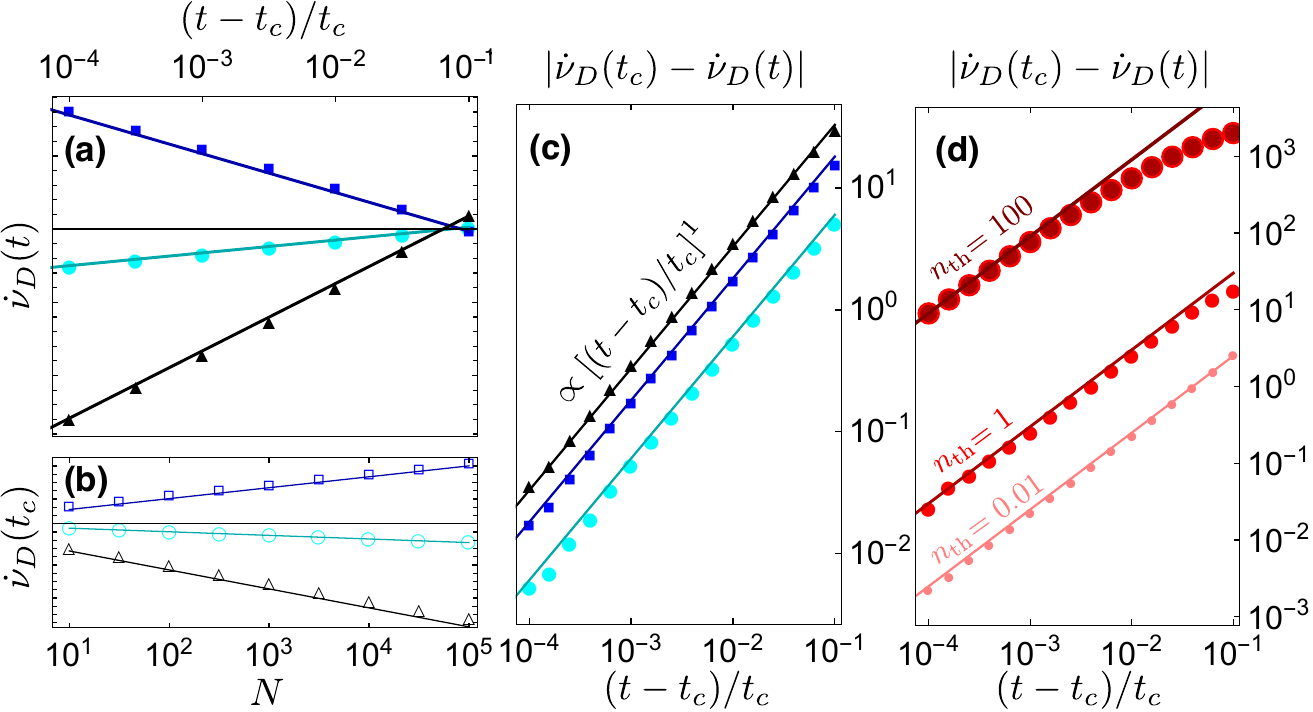}
\caption{%
Scaling of $\dot\nu_D$:
In (a) and (b) logarithmic scaling of the order parameter is shown for $t_c=\{0.5,1.5,2.5\}\overline\tau$ with cyan circles, blue squares, and black triangles, respectively. The lines are logarithmic fits with Eq.~(\ref{scal1}) in (a) and Eq.~(\ref{scal2}) in (b).
(c) The power law scaling for the three first integer multiple critical times is presented. The lines are power law fits $\propto(t-t_c)$ in each case.
(d) The power law scaling at $t_c=\overline\tau$ for three finite temperatures in units of fundamental mode occupation number $n_{\rm th}$. The lines are linear fits.
In all plots the system size is $N=10^5$.
}
\label{fig:impel}%
\end{figure}
%

%
%
%========================================%
\textit{Pure dephasing.---}
Let us now study decay of the coherence in the qubit when initially prepared in a superposition state. The off-diagonal elements of the spin density matrix $\varrho$ at every instance of time evolve as
$\bra{\uparrow}\varrho(t)\ket{\downarrow} = \bra{\downarrow}\varrho(t)\ket{\uparrow} = \bra{\uparrow}\varrho(0)\ket{\downarrow}e^{\Gamma(t)}$
with~\cite{Breuer2007}
\begin{equation}
\Gamma(t)=-\sum_{k=0}^N \big(\frac{g_k}{\omega_k}\big)^2 \coth\!\big(\frac{\hbar\omega_k}{2k_{\rm B} T}\big)(1-\cos\omega_k t),
\label{deph}
\end{equation}
for a thermal bosonic bath at temperature $T$, while $k_{\rm B}$ is the Boltzmann constant.
For a zero-temperature bath $T=0$ it is enough to drop the $\coth$ factor.
One then obviously finds that the free-induction decay (FID) of the spin-qubit is the same as the Loschmidt amplitude of the bosonic bath when quenched from a non-displaced form to a `properly' displaced bath: $\mathcal{G}(t)=\exp\{\Gamma(t)\}$.
It is noteworthy to mention that this is equivalent to the generalized outcome of a Ramsey measurement~\cite{Asadian2014}.
The FID is plotted in Fig.~\ref{fig:fid}(a). The coherence of the qubit falls off with time. However, close to global period of the bath, $\tau_0$, it sets for a partial revival depending on the strength of the qubit-bath coupling $g_0$.
Such revivals signal non-Markovian nature of the qubit dynamics. More interestingly, kinks and cusps are observable during the collapse and revivals around half of the collective periods $\overline\tau \equiv 2\pi/\delta$. Such behavior stems from DTPT in the bosonic bath manifesting itself in the coherence dynamics of the spin.
The numerically computed Fisher zeros indeed confirm this observation. One clearly sees from Fig.~\ref{fig:fid}(c) that the Fisher zeros exhibit tongues appearing exactly at the integer multiples of $\overline\tau/2$.
And as the size of system increases the zeros of the Loschmidt amplitude get closer to the real axis and the tongues are expected to cross the real time axis.

We next study the DTPTs in our system both qualitatively and quantitatively. For this, the order parameter $\nu_D$ and its first derivative are numerically evaluated and plotted in Fig.~\ref{fig:fid}(b). One notices that kinks at the half integer multiples of $\overline\tau$ are of logarithmic transitions, while at the integer multiples of the period the system undergoes transitions of type (ii).
In order to quantify them, we study the scaling behavior of the cusps with the two different methods described above.
Regarding the logarithmic divergences; we find that for the first three critical times namely $t_c=\{0.5,1.5,2.5\}\overline\tau$ the extremum values scale logarithmically as in Eq.~(\ref{scal1}) with the coefficients $\kappa\approx\{-3.8,11.5,-20\}$, respectively. Meanwhile, the critical behavior of $\dot\nu_D$ at these points follows the function in Eq.~(\ref{scal2}) with $\kappa'\approx\{3.8,-11.5,20\}$ [Fig.~\ref{fig:impel}(a) and (b)]. Therefore, the dynamical universal exponent of the system equals unity. The same value is computed for the kinks at integer multiples of $\overline\tau$ when we fit a power law function at these points [Fig.~\ref{fig:impel}(c)].

%
%
%========================================%
%\textit{Finite temperature robustness.---}
In order to examine possibility of tracing such phase transition in a real world experiment, we take into account the effect of finite temperature. The results are summarized in Fig.~\ref{fig:impel}(d) where we plot the scaling behavior at $t_c=\overline\tau$ for three different temperatures in units of fundamental mode occupation number $n_{\rm th}=\big[\exp\{\hbar\omega_0/k_{\rm B}T\} -1 \big]^{-1}$.
The exact calculations show that the kinks and cusps in the FID and derivative of the order parameter survive at sufficiently low temperatures. However, as the temperature rises, the kinks start to turn into smooth edges as do the jumps in $\dot\nu_D$ (see Supplemental Material for the plots~\cite{suppinfo}). In fact, deviations from the power law scaling set off as the temperature rises the fundamental mode occupation number to $n_{\rm th}\gtrsim 1$. But the power law scaling remains detectable even at temperatures as high as $n_{\rm th}\sim 100$.
For a membrane with fundamental frequency $\omega_0/2\pi = 20$~MHz this occupation number corresponds to an ambient temperature of $T\approx 0.1$~K. At temperatures as low as $T=15$~mK the number reduces to $n_{\rm th}\approx 15$, therefore, making observation of the effect even more accessible.

%
%
%----------ACKNOWLEDGEMENT----------%
\begin{acknowledgements}
%\textit{Acknowledgements.---}%
The author thanks R. Jafari, A. Asadian, and F. Shahbazi for helpful discussions.
The support by STDPO and IUT through SBNHPCC is acknowledged.
\end{acknowledgements}

%
%
%----------REFERENCES----------%
\bibliography{DTPT}

\end{document}